# Petahertz optical response in graphene

Matthias Baudisch[1], Andrea Marini[1], Joel D. Cox[1], Tony Zhu[2], Francisco Silva[1], Stephan Teichmann[1], Mathieu Massicotte[1], Frank Koppens[1,3], Leonid S. Levitov[2], F. Javier García de Abajo[1,3], Jens Biegert[1,3]

1 ICFO - Institut de Ciencies Fotoniques, The Barcelona Institute of Science and Technology, 08860 Castelldefels (Barcelona), Spain
2 Department of Physics, Massachusetts Institute of Technology, Cambridge, MA 02139, USA
3 ICREA, Pg. Lluís Companys 23, 08010 Barcelona, Spain

e-mail: jens.biegert@icfo.eu

**The temporal dynamics of charge carriers determines the speed with which electronics can be realized in condensed matter, and their direct manipulation with optical fields promises electronic processing at unprecedented petahertz frequencies (1), consisting in a million-fold increase from state of the art technology. Graphene (2) is of particular interest for the implementation of petahertz optoelectronics due to its unique transport properties, such as high carrier mobility (3) with near-ballistic transport (4) and exceptionally strong coupling to optical fields (5–7). The back action of carriers in response to an optical field is therefore of key importance towards applications. Here we investigate the instantaneous response of graphene to petahertz optical fields and elucidate the role of hot carriers on a sub-100 fs timescale. Measurements of the nonlinear response and its dependence on interaction time and field polarization allow us to identify the back action of hot carriers over timescales that are commensurate with the optical field. An intuitive picture is given for the carrier trajectories in response to the optical-field polarization state. We note that the peculiar interplay between optical fields and charge carriers in graphene may also apply to surface states in topological insulators (8) with similar Dirac cone dispersion relations.**

Graphene is a remarkable material (*2*) that exhibits fascinating properties such as ultrahigh electron mobility (*3, 4*), large mechanical strength (*9*), and extraordinary optoelectronic behaviour (*10–12*). In particular, its electron transport properties (*13, 14*), which are described by a two-dimensional gas of massless Dirac fermions moving at a velocity of 1 nm/fs, are of interest to develop novel optoelectronic devices. Its room-temperature carrier mobility of up to $1.5 \times 10^6$ cm$^2$/Vs is unusually high (*3*), while the third-order susceptibility of a single-atom-thick graphene layer is $10^8$ times stronger than in a dielectric material (*5*). The large nonlinear response of graphene (*5, 7, 15, 16*) is attributed to its linear electronic energy dispersion (17), which not only provides resonant interband transitions for a continuous range of low photon energies, but in the presence of intense optical fields leads to square-wave oscillatory motion of Dirac fermions (i.e. an anharmonic current response of the material and the emission of new frequencies of light). Interestingly, while third-harmonic generation (*17, 18*) and four-wave mixing were observed early on, there is an unexpected absence of reports on higher-order nonlinear optical phenomena in the literature, despite attempts involving high optical field strengths (*19*). However, an extreme nonlinear effect has been recently reported: optical pumping can dramatically affect the optical response and produce transient plasmons (*20*), which have been speculated to lead to order-unity optically-induced modulation of absorption in patterned monolayer graphene (*21*), further increasing light-matter interactions. Moreover, the introduction of a slight band gap in the graphene dispersion effectively enhances the nonlinear response compared to that of gapless graphene (*22, 23*). Consequently, the nontrivial interactions among optical fields and Dirac fermions in graphene, even at low field strengths, demands an investigation of the non-equilibrium response of the material's charge carriers as a first step toward the implementation of petahertz optoelectronic control (*1*).

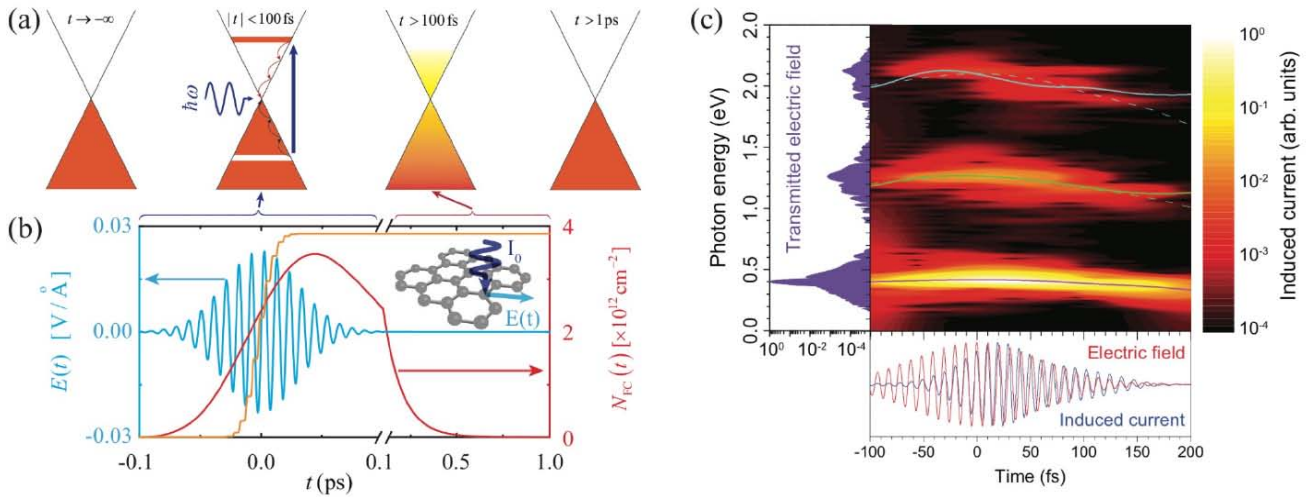

**Figure 1: Nonlinear temporal carrier dynamics in graphene.** (a) Schematic representation of the ultrafast temporal dynamics of photo-excited electrons in extended graphene. (b) Temporal evolution of free carriers (red curve, right vertical axis) generated by a normally-impinging linearly-polarized Gaussian optical pulse (cyan curve, left axis) of 7 GW/cm$^2$ peak intensity and 50 fs FWHM duration, as obtained from the MDF model (see Methods) for monolayer graphene. A characteristic ionized fraction of an atomic gas target is shown for comparison as calculated with PPT theory (*24*) (orange curve). (c) Gabor analysis (150 fs Gaussian time window) of the squared induced current (color plot), along with the spectral (left plot) and temporal (lower plot) dependences of the transmitted electric field. Solid curves superimposed onto the color plot correspond to the time-dependent spectral centroid of the fundamental and harmonic intensities, while the two dashed curves are obtained by multiplying the energy of the fundamental centroid by factors of 3 and 5.

Figure 1a illustrates the temporal dynamics of graphene electrons upon intense optical pulse irradiation over various timescales. In contrast to other materials, the conical band structure of graphene facilitates the excitation of electron-hole pairs at any instant of time during irradiation with an optical field, thus resulting in a near-instantaneous non-thermal electronic response of free charge carriers. This is followed by electronic thermalization due to electron-electron collisions over a characteristic time scale of ~50 fs (*25*, *26*). Finally, electron-phonon coupling leads to electron relaxation and electron-hole recombination, thereby gradually reducing the electron temperature over a characteristic picosecond time scale. The described non-equilibrium carrier dynamics of graphene have been experimentally investigated, in particular through measurements in the THz regime (*25*, *27*, *28*) and also using ca. 10-fs-duration pulses in the visible (*26*). Such measurements were instrumental to elucidate the dynamics of charge carriers in this material, but an implementation of ultrafast optoelectronics requires additional knowledge on how charge carriers influence the optical fields themselves.

Here, we investigate the direct back-action of the carrier dynamics on the optical field at the petahertz scale through harmonic generation. Direct back-action is important because ultrafast petahertz-scale pulses carry significant spectral bandwidth and their interaction involves substantial reshaping and frequency modulation of the optical field as well as of the electron distribution. We employ two different theoretical models to describe this interaction (see Methods and SI): one based on a non-perturbative continuum picture in which π electrons are treated as massless Dirac fermions (MDFs); and complementary time-domain simulations combining an atomistic tight-binding model for the electrons with the random-phase approximation formalism (TB-RPA). We find that these two approaches yield results in excellent agreement for the experimentally relevant conditions.

Figure 1b shows the predicted temporal evolution of free carriers (red curve) in response to the optical field (blue curve) of an experimental pulse with 70-fs FWHM duration at a photon energy of 0.4 eV and with peak intensity of 5 GW/cm$^2$. The absence of a band gap, in combination with the linear energy dispersion of graphene, leads to an instantaneous response of free carriers to the optical field. This behavior is markedly different compared to systems with a large band gap, such as free atoms and insulators (*1*, *29*, *30*). To illustrate the difference, we show the temporal evolution of the ionization yield for an atomic medium (orange curve), where the well-known step-like response of the emitted electrons is clearly discernable. Depending on the strength of the optical field, the medium may be fully ionized already at the leading edge of the pulse, in such a way that the trailing edge experiences a constant carrier density. The scenario is distinctly different in graphene because free carriers are instantaneously generated whenever an optical field is present. In addition, efficient and rapid electron-electron scattering occurs due to the high mobility of free carriers, thus affecting the optical field in a transient manner, which is markedly different from the abrupt temporal variation of the atomic medium.

Consequently, in graphene the generated free carriers oscillate synchronously with the optical field at early times (leading edge of the pulse), but modify the petahertz optical response for later times during the optical field (trailing part of the pulse). The result is an induced asymmetry on the leading and trailing parts of the optical field (Fig. 1b), and a switchover from dielectric- to metallic-like material response along the evolution of the optical field. The bottom panel in Fig. 1c shows the resulting temporal response of the simulated induced current, which confirms that free carriers oscillate synchronously with the optical field at early times, but later acquire a phase lag due to the increasing density of free carriers.

The back action of charge carrier motion to the optical field is revealed by the time-energy plot of the resulting petahertz light emission in Fig. 1c. New optical frequencies, at odd orders of the fundamental (left panel), are generated due to the anharmonic response from the optical-field-driven square-wave oscillatory motion of Dirac fermions. Additionally, the free carriers lower the refractive index and cause a spectral blue shift that is proportional to the optical frequency (main panel). Because of this interplay, the optical pulse accelerates at the trailing edge, while the pulse envelope undergoes self-steepening at the leading front.

Experimentally, direct observation of back action in the temporal domain requires petahertz-level sub-cycle sampling of the optical field, but the effect is clearly revealed in the spectral domain through blue shifts and the generation of new optical frequencies. For our experiment, we used 70-fs (6.4 optical cycles) FWHM pulses with a central wavelength of 3.1 µm (0.4 eV photon energy) from a mid-IR OPCPA (*31*). We focused the pulses at normal incidence onto a 5-monolayer graphene sample that was transferred onto a 0.4-mm thick CaF$_2$ substrate (see Fig. 2a and Methods for details). Reference measurements were performed on the CaF$_2$ substrate without graphene to exclude any additional contribution for peak intensities below 10 GW/cm$^2$. Graphene irradiation at higher peak intensity resulted in an irreversibly decreasing signal over time, indicating material damage. The transmitted mid-IR field and the generated new optical frequencies were collected with a spectrometer.

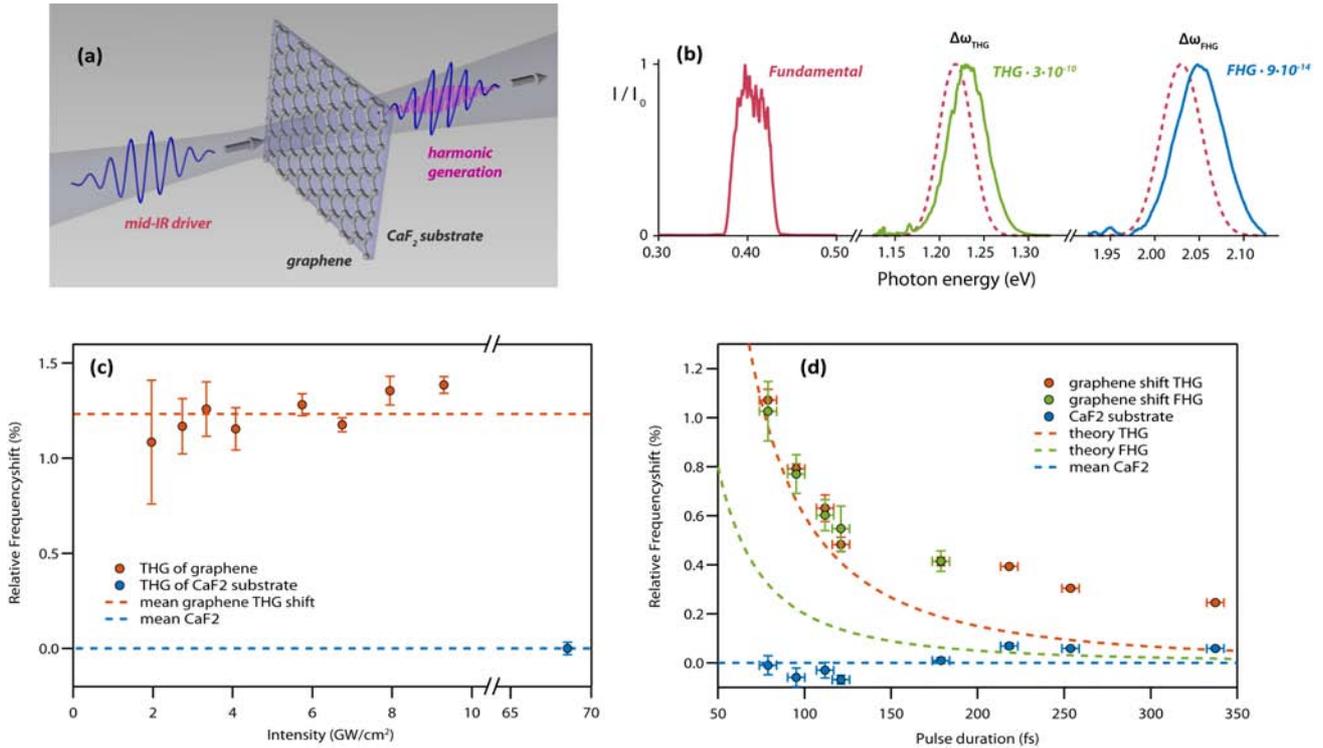

**Figure 2: Optical response to carrier dynamics in graphene.** (a) Schematic of the experimental configuration, showing the mid-IR field (linearly polarized, 70-fs pulses at 3.1 μm wavelength, i.e., $E_0$=0.4 eV photon energy) before and after propagation through 5 monolayers of graphene that are supported on a 0.4 mm thick $CaF_2$ substrate. Pulses of different durations and degrees of elliptical polarization are also investigated. (b) Measured fundamental spectrum, along with the emission at the third and fifth harmonics (blue-shifted). The dashed curves represent the nominal position of the third and fifth harmonic. (c) The resulting third harmonic is blue shifted by 1.8% from $3E_0$ and the blue shift is independent of driving field intensity. (d) The blue shift depends inversely on pulse duration for both the third and the fifth harmonic.

A characteristic measurement for linear polarization is shown in Fig. 2b for irradiation at a peak intensity of 5 GW/cm². We observed the generation of harmonic frequencies up to fifth order and with proportional blue-shifts compared to the nominal third and fifth harmonic (dashed curves). This first observation of the fifth harmonic at petahertz frequencies is commensurate with predictions (*7, 15, 32, 33*) and contradicts earlier measurements in multilayer graphene at terahertz frequencies in which the absence of optical harmonics was attributed to fast carrier scattering (*19*). To further investigate the influence of the charge carriers, we measured the scaling of the blue shift with peak intensity and pulse duration. Longer pulses, up to 340 fs duration, at constant peak intensity were generated by spectrally narrowing the OPCPA's output and increasing the pulse energy proportionally with the increase of pulse duration. We resorted to spectral narrowing rather than simple frequency chirping to exclude additional dynamical effects associated with pulse propagation in graphene. Figure 2c shows that the measured blue-shifts did not depend on peak intensity for constant pulse duration. By fixing the peak intensity and changing the duration of the pulse we observed however that the blue shifts of the third and fifth harmonics scaled inversely with pulse duration. These observations are in excellent qualitative agreement with theoretical simulations, which are overlaid in Figs. 2c and d (dashed curves).

To gain further insight into carrier dynamics and the possible influence of the k-space on the interaction, we repeated our measurements with varying polarization for the incident mid-IR field. Figure 3a shows that the third harmonic signal strength decreased rapidly with increasing ellipticity $\epsilon$ and that the resulting third harmonic was two orders of magnitude weaker for a circularly polarized field ($\epsilon = 1$) compared with a linearly polarized field ($\epsilon = 0$). Of particular interest is the dependence of the blue shift on ellipticity: Figure 3b shows that the blue shift was maximal (approximately 2% of the central frequency) for linear fields and decreased to values lower than 0.3% for circularly polarized light.

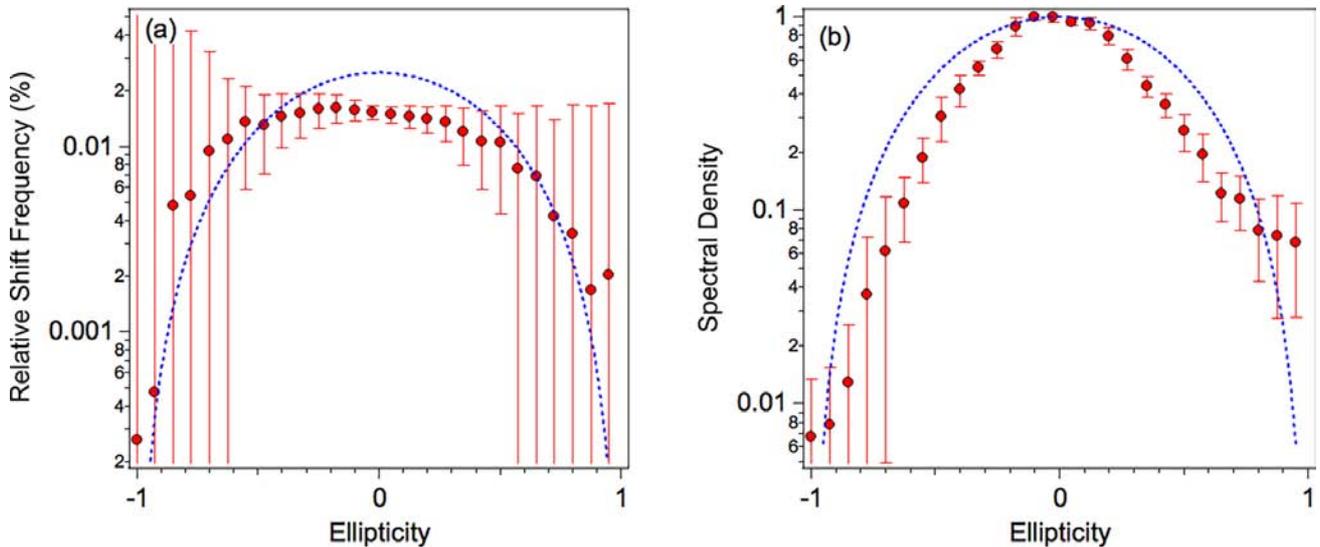

**Figure 3: Vectorial field-driven carrier dynamics.** We show the dependence of the third-harmonic blue shift (a) and intensity (b) as a function of ellipticity of the impinging light. Measurements (symbols with error bars) are compared with MDF simulations (solid curves).

Our simulations qualitatively reproduce the experimental findings and are indicated by the solid blue lines superimposed on the experimental data presented in Figs. 3. It should be noted that, while the reduction of generated free carriers in atoms and molecules due to the scaling of ionization and reduction of the refractive index (*34*) with circular polarization are well-known phenomena, the scenario here is markedly different due to the availability of carriers, the ballistic transport of Dirac fermions and the band topology in graphene. Our simulations clearly confirm a reduction of anharmonic response, and hence a diminishing harmonic generation, and they reproduce the measured reduction in blue shift with increasing ellipticity. The origin of this behaviour becomes clear from a trajectory analysis of field-driven Dirac fermions, as shown in Fig. 4 for the extreme cases of linear and circularly polarized driving fields. With linear polarization, low-energy carriers are field-driven across the anharmonic potential of the Dirac cone (Fig. 4a) and the emission of harmonics is the direct consequence of such anharmonic motion. During their excursion in the Dirac potential, there is a large probability that the charge carriers re-encounter the lowest energy point of the Dirac cone, where electron scattering leads to a significant generation of free carriers and a consequential blue shift. The dynamics are very different when graphene is driven by circularly-polarized light (Fig. 4b), for which charge carriers follow spiralling trajectories along the Dirac cone. In contrast to the linear driving scenario, the carriers do not encounter the lowest energy point twice during each field cycle, but instead only at the starting and ending parts of the pulse. More precisely, under the action of a circularly-polarized pulse, carriers are driven in spiralling trajectories ascending the Dirac cone with increasing field amplitude and back down with diminishing field amplitude, thus acquiring a $2\pi$ phase in k-space during each optical cycle. The availability of carriers across k-space ensures that a matching trajectory always exists that is $\pi$ out of phase, so that any anharmonic response is effectively cancelled for circular driving fields.

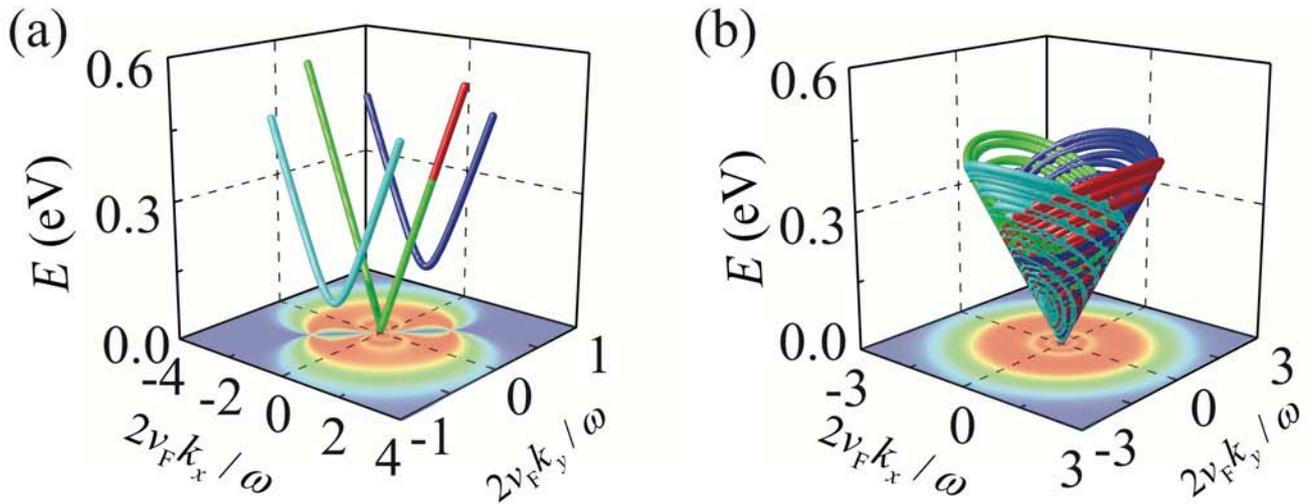

**Figure 4: Trajectories of photo-excited carriers in graphene.** We show trajectories for linear (a) and circularly polarized (b) incident light and different initial k-space conditions (plotted with different colors, see SI for more details). The out-of-equilibrium k-space density of free carriers is shown in the lower color plot.

In conclusion, we identify both the dynamics of graphene Dirac carriers driven by optical petahertz fields and the back action of these dynamics on the driving field as revealed by blue shifts in the generation of third and fifth harmonics. Theoretical modelling explains the results and allows us to track down carrier trajectories in the specific topology of Dirac cones, which readily explains the disappearance of the blue shifts when the incident light is circularly polarized. These investigations of the dynamics of Dirac carriers in graphene provide important insight for the understanding and design of future graphene-based ultrafast optoelectronic devices, while they open an avenue toward to exploration of topology-dependent carrier dynamics in other 2D systems, such as surface states of topological insulators and other 2D materials.

## METHODS SECTION

*Theoretical Models:* Our simulations are based on two complementary approaches to describe the nonlinear optical response of graphene: (i) a non-perturbative continuum picture in which π electrons are treated as massless Dirac fermions (MDFs); and (i) an atomistic tight-binding model for the electrons combined with the random-phase approximation formalism (TB-RPA). These two approaches yield results in excellent mutual agreement under the conditions of the present experiment. (i) The MDF model [13] introduces light-matter interaction through the electron quasi-momentum $\mathbf{k} + (e/c)\mathbf{A}$, where $\mathbf{k}$ is the electron momentum, whereas $\mathbf{A}(t) = -c \int_{-\infty}^{t} \mathbf{E}(t') \, dt'$ is obtained from the incident light electric field $\mathbf{E}$. Electrons are treated independently and their dynamics is described by the Dirac equation for massless fermions, which can be recast as in the form of Bloch equations (32, 35, 36), involving the population inversion $n_\mathbf{k}(\mathbf{R}, t)$, the interband coherence $\rho_\mathbf{k}(\mathbf{R},t)$ (35), and a phenomenological relaxation time τ. We numerically integrate these equations to extract the $\mathbf{k}$-resolved induced current and is integral over $\mathbf{k}$ to yield the total current. A self-contained presentation of this method is given in the SI, along with analytical expressions for the $\mathbf{k}$-space resolved contribution to harmonic generation under CW illumination. Pulses are also simulated by direct time integration of the Bloch equations. (ii) In the TB-RPA method (37, 38), we solve the one-particle density matrix $\dot{\rho} = -\left(\frac{i}{\hbar}\right)[H_{TB} - e\phi, \rho] - \left(\frac{1}{2\tau}\right)(\rho - \rho^\circ)$ to describe the temporal dynamics of graphene electrons, where $H_{TB}$ is the nearest-neighbours tight-binding Hamiltonian (2.8 eV hopping energy, one orbital per carbon site) and ϕ is the electric potential produced by the incident light plus the Hartree interaction. We express the density matrix $\rho = \sum_{jj'} |j\rangle \langle j'|$ in terms of the one-electron eigenstates of $H_{TB}$ with complex-number expansion coefficients $\rho_{jj'}$. The relaxed state of the system is $\rho^\circ_{jj'} = \delta_{jj'}$ for occupied states and 0 otherwise. The spectrally resolved induced field is then readily obtained by Fourier transforming the $\rho_{jj'}$ coefficients (38). Calculations are performed for undoped ribbons of increasing width until converge is achieved. We assume a relaxation time $\tau$ = 100 fs in both approaches.

# REFERENCES

1. M. Garg *et al.*, Multi-petahertz electronic metrology. *Nature*. **538**, 359–363 (2016).
2. K. S. Novoselov *et al.*, A roadmap for graphene. *Nature*. **490**, 192–200 (2012).
3. L. Wang *et al.*, One-Dimensional Electrical Contact to a Two-Dimensional Material. *Science* **342**, 614–617 (2013).
4. L. Banszerus *et al.*, Ballistic Transport Exceeding 28 μm in CVD Grown Graphene. *Nano Lett.* **16**, 1387–1391 (2016).
5. E. Hendry, P. J. Hale, J. Moger, A. K. Savchenko, S. A. Mikhailov, Coherent Nonlinear Optical Response of Graphene. *Phys. Rev. Lett.* **105**, 97401 (2010).
6. S. A. Mikhailov *et al.*, Non-linear electromagnetic response of graphene. *Europhys. Lett.* **79**, 27002 (2007).
7. A. R. Wright, X. G. Xu, J. C. Cao, C. Zhang, Strong nonlinear optical response of graphene in the terahertz regime. *Appl. Phys. Lett.* **95**, 72101 (2009).
8. M. Z. Hasan, C. L. Kane, *Colloquium* : Topological insulators. *Rev. Mod. Phys.* **82**, 3045–3067 (2010).
9. C. Lee, X. Wei, J. W. Kysar, J. Hone, Measurement of the Elastic Properties and Intrinsic Strength of Monolayer Graphene. *Science* **321** (2008).
10. J. Wang, Y. Hernandez, M. Lotya, J. N. Coleman, W. J. Blau, Broadband Nonlinear Optical Response of Graphene Dispersions. *Adv. Mater.* **21**, 2430–2435 (2009).
11. Z. Fei *et al.*, Infrared Nanoscopy of Dirac Plasmons at the Graphene–$SiO_2$ Interface. *Nano Lett.* **11**, 4701–4705 (2011).
12. A. E. Nikolaenko *et al.*, Nonlinear graphene metamaterial. *Appl. Phys. Lett.* **100**, 181109 (2012).
13. K. I. Bolotin *et al.*, "Ultrahigh electron mobility in suspended graphene" *Sol. St. Commun.* **146**, 351-355 (2008).
14. K. Chen *et al.*, Diversity of ultrafast hot-carrier-induced dynamics and striking sub-femtosecond hot-carrier scattering times in graphene. *Carbon N. Y.* **72**, 402–409 (2014).
15. S. A. Mikhailov *et al.*, Nonlinear electromagnetic response of graphene: frequency multiplication and the self-consistent-field effects. *J. Phys. Condens. Matter*. **20**, 384204 (2008).
16. H. Rostami, M. I. Katsnelson, M. Polini, Theory of plasmonic effects in nonlinear optics: The case of graphene. *Phys. Rev. B*. **95**, 35416 (2017).
17. N. Kumar *et al.*, Third harmonic generation in graphene and few-layer graphite films. *Phys. Rev. B*. **87**, 121406 (2013).
18. S.-Y. Hong *et al.*, Optical Third-Harmonic Generation in Graphene. *Phys. Rev. X*. **3**, 21014 (2013).
19. M. J. Paul *et al.*, High-field terahertz response of graphene. *New J. Phys.* **15**, 85019 (2013).
20. G. X. Ni *et al.*, Ultrafast optical switching of infrared plasmon polaritons in high-mobility graphene. *Nat. Photonics*. **10**, 244–247 (2016).
21. F. J. García de Abajo, Graphene Plasmonics: Challenges and Opportunities. *ACS Photonics*. **1**, 135–152 (2014).
22. Y. S. Ang, C. Zhang, Subgap optical conductivity in semihydrogenated graphene. *Appl. Phys. Lett.* **98**, 42107 (2011).
23. Y. S. Ang, Q. Chen, C. Zhang, Nonlinear optical response of graphene in terahertz and near-infrared frequency regime. *Front. Optoelectron.* **8**, 3–26 (2015).
24. P. V. S. and T. M. V Perelemov A M, Ionization of atoms in an alternating electric field. *Sov. Phys. JETP*. **23**, 1393–1409 (1966).
25. I. Gierz *et al.*, Snapshots of non-equilibrium Dirac carrier distributions in graphene. *Nat. Mater.* **12**, 1119–1124 (2013).
26. D. Brida *et al.*, Ultrafast collinear scattering and carrier multiplication in graphene. *Nat. Commun.* **4**, 611–622 (2013).
27. P. Bowlan, E. Martinez-Moreno, K. Reimann, T. Elsaesser, M. Woerner, Ultrafast terahertz response of multilayer graphene in the nonperturbative regime. *Phys. Rev. B*. **89**, 41408 (2014).
28. Z. Mics *et al.*, Thermodynamic picture of ultrafast charge transport in graphene. *Nat. Commun.* **6**,

7655 (2015).
29. A. Schiffrin *et al.*, Optical-field-induced current in dielectrics. *Nature*. **493**, 70–74 (2012).
30. M. Schultze *et al.*, Controlling dielectrics with the electric field of light. *Nature*. **493**, 75–78 (2012).
31. M. Baudisch, B. Wolter, M. Pullen, M. Hemmer, J. Biegert, High power multi-color OPCPA source with simultaneous femtosecond deep-UV to mid-IR outputs. *Opt. Lett.* **41**, 3583 (2016).
32. K. L. Ishikawa, Nonlinear optical response of graphene in time domain. *Phys. Rev. B*. **82**, 201402 (2010).
33. S. Shareef, Y. S. Ang, C. Zhang, Room-temperature strong terahertz photon mixing in graphene. *J. Opt. Soc. Am. B*. **29**, 274 (2012).
34. R. W. Boyd, *Nonlinear optics* (Academic Press, 2003).
35. A. Marini, J. D. Cox, F. J. G. de Abajo, Theory of graphene saturable absorption. *Phys. Rev. B* **95**, **125408** (2016).
36. I. Al-Naib, J. E. Sipe, M. M. Dignam, High harmonic generation in undoped graphene: Interplay of inter- and intraband dynamics. *Phys. Rev. B*. **90**, 245423 (2014).
37. S. Thongrattanasiri, A. Manjavacas, F. J. García de Abajo, Quantum Finite-Size Effects in Graphene Plasmons. *ACS Nano*. **6**, 1766–1775 (2012).
38. J. D. Cox *et al.*, Electrically tunable nonlinear plasmonics in graphene nanoislands. *Nat. Commun.* **5**, 5725 (2014).

**ACKNOWLEDGEMENT**

We thank D. Zalvidea, for helpful and inspiring discussions. We acknowledge financial support from the Spanish Ministry of Economy and Competitiveness (MINECO), through the "Severo Ochoa" Programme for Centres of Excellence in R&D (SEV-2015- 0522) and the grants FIS2014-56774-R, FIS2014-51478-ERC and MAT-2014-59096; the Catalan Institució Catalana de Recerca I Estudis Avançats; Agencia de Gestió d'Ajuts Universitaris i de Recerca (AGAUR) with grant SGR 2014-2016; the Fundació Cellex Barcelona; the European Union's Horizon 2020 research and innovation program under LASERLAB-EUROPE (EU-H2020 654148); COST Actions MP1203, XUV/X-ray light and fast ions for ultrafast chemistry (XLIC); the Marie Sklodowska-Curie grant agreement 641272.

**Author Contribution**

The experiment was devised by J.B. M.B. conducted the experiment and analysed the data with help from J.B. F.S. and S.T. M.M. and F.K. provided the samples. A.M., J.D.C. and J.G.d.A. provided modelling and theoretical analysis. T.Z. and L.S.L. provided additional theoretical modelling. The manuscript was written by A.M., J.G.d.A. and J.B. All authors discussed and commented on the manuscript.